\documentstyle[aps,prd,psfig]{revtex}\voffset=.5truein
\begin{document}\title{Revising neutrino oscillation parameter space with
direct flavor-changing interactions}
\author{ Loretta M. Johnson$^a$ and Douglas W. McKay$^b$}
\address{$^a$Department of Physics, Grinnell College, Grinnell, IA 50112; 
$^b$Department of Physics and Astronomy, University of Kansas, Lawrence,
KS 66045\\} \maketitle

\begin{abstract} We formulate direct, neutrino flavor-changing
interactions in a framework that fits smoothly with the parameterization
of two- and three-state mixing of massive neutrino states. We show that
even small direct interaction strengths could have important consequences
for the interpretation of currently running and proposed oscillation
experiments. The oscillation amplitude and the borders of the allowed
regions in two- and three-flavor mixing parameter space can be sensitive to
the presence of direct interactions when the transition probability is small.
We use extensively the high sensitivity of the NOMAD experiment to illustrate
potentially large effects from small, direct flavor violation. In the purely
leptonic sector, we find that the clean $\nu_{\mu}$ and $\nu_{e}$ beams
from a $\mu^+ - \mu^-$ collider could provide the sharpest tests of direct
flavor violation. 
 \end{abstract}

\section{Introduction}
The  Superkamiokande collaboration's zenith angle analysis of its data
 \cite{SKPRL}, may sway 
even a skeptic to the view that neutrino oscillations have been observed. 
The precise form of the
neutrino mass matrix and the number of neutrino species remain undetermined, 
however. The confusing and complicated nature of the whole
collection of neutrino data in laboratory and astrophysical settings makes for
an  exciting playing field for workers trying to  
establish a ``standard model'' of neutrino physics.  Vigorous efforts have
been made to determine whether one needs only three flavors of neutrinos,
mixed either two-by-two \cite{Lang1} or with some form of full three-
state mixing\cite{cf96},\cite{swede} or whether a fourth, sterile neutrino
is needed in the mixture \cite{bpww1},\cite{Lang2}.
There is not yet a compelling phenomenological or theoretical case
for any of these reasonable, and partially successful, approaches.  

Direct violation in neutrino interactions is a feature of the flavor-mixing 
puzzle that has received relatively little attention.  Most models that
predict the existence of neutrino mass and mixing, and
consequently neutrino oscillations, also contain direct neutrino flavor-
violating interactions.  In fact, some interesting models have no neutrino
masses at tree level, but flavor-violating interactions generate them in
loop graphs\cite{mm79},\cite{az80},\cite{babu},\cite{gaur},\cite{jv1197}.
Moreover, with the help of resonant enhancement, massless neutrinos
can mix in a non-trivial way in certain models and produce flavor transitions 
in flight\cite{ww83}.  It seems natural to extend the phenomenological
framework for oscillations of propagating neutrinos to include the effects 
of direct interactions in a way that allows one to survey all effects at once.
A formalism to carry out this program is sketched and illustrated
in\cite{jm98}, and we extend the range of applications and uncover several new 
features in the present paper. Specific models are not discussed here, since
we emphasize the model independent features of the oscillation-plus-direct
flavor violation analysis.  Bounds on parameter combinations from a given
model can be directly obtained from the coefficients of our effective four
Fermi Lagrangian.

As in \cite{jm98}, our primary concern is with
the accelerator experiments \cite{ladar}\cite{ladif}\cite{krdar}.  The
experimental constraints on muonium-antimuonium conversion are now so tight
\cite{mumu} that a purely direct
interaction explanation of the muon decay-at-rest (DAR) signal reported
in \cite{ladar} can be rather convincingly ruled out\cite{BG}.  Here we
emphasize  effects that are significant when combined with the
oscillation phenomenon.  For example, we show in section IV that direct effects
can kill oscillations in special circumstances.  We also find that in
high sensitivity experiments, where tight limits in regions of
$sin^2 2\theta$-$\Delta m^2$ space are  achieved, small, direct flavor
violation
can change a boundary by more than an order of magnitude.  We illustrate this
by showing examples of the effects on the large $\Delta m^2$, small
$sin^2 2\theta$ boundary of the $\nu_{\mu}\leftrightarrow\nu_{\tau}$ mixing
set by the NOMAD collaboration\cite{NOM}. In another sensitive comparison,
we show the power of comparing the ``wrong flavor'' appearance signals from
the clean $\nu_{\mu}$ and $\nu_{e}$ beams afforded by a $\mu^+ - \mu^-$ 
collider.  

There are several studies where direct flavor violation is considered in
the solar and atmospheric cases.  In \cite{berg} the combined oscillation and
direct effects are applied to an analysis of the resonant conversion of 
electron neutrinos to other species as the explanation of the solar neutrino 
deficit\cite{davis}\cite{gallex}\cite{sage}\cite{kksol}, while in \cite{GG} 
and \cite{forn}, an explanation using resonantly enhanced direct lepton number 
violation of the zenith angle effect reported in \cite{SKPRL}
was presented.  In both cases significant effects were reported,
though a complete explanation in terms of direct interactions is probably
not possible in the solar case.  The situation is not settled in the
atmospheric case.  A critique of alternatives to the large $\nu_{\mu}$-
$\nu_{e}$ mixing  solutions to the atmospheric
neutrino data such as those presented in \cite{GG},\cite{forn}
and \cite{vbdec} is given in \cite{LL}.  A model using $\nu_{\mu}$ decay that
answers the objection in \cite{LL} is described in  \cite{vbdec2}.
 
We encourage the reader interested in the impact of small, direct flavor
violation on the analysis of experiments with high sensitivity to `` wrong
flavor'' appearance to go straight to Sec. 4.  The background is given in the
next two sections.
In the following section, the parameterization and notation are defined, and
the notion of a generalized transition probability factor is explained.  In
Sec. 3 the experimental constraints on flavor violating parameters defined
in Sec. 2 are summarized.  As just remarked,
the formalism is applied to a number of examples
drawn from current or future accelerator experiments in Sec. 4.  We keep
the constraints on flavor-violating parameters clearly in mind in the 
discussion of these applications.  The formalism and results of its 
applications are summarized and several conclusions are drawn in Secs. 5
and 6.  The general forms of the probability factors that apply to the 
case where $\mu$ decay provides the source of neutrinos are given in the
Appendix.  
\medskip

\section{Formalism}
\medskip
In this section we develop a compact parameterization of direct interaction
effects in neutrino flavor-changing processes.  We represent the
low-energy effective interactions involving neutrinos, charged leptons and
first generation quarks by the four-Fermi semileptonic (S) and leptonic (L)
Lagrangians\footnote{The CKM factors multiplying $G_{F}$ play no direct role
in our discussion, so they are suppressed in the notation.}

\begin{equation} \-$\it{$\mathcal{L}$}$^S=2\sqrt{2}G_{F}K^h_{Aij}(\overline
{l_{i}}\Gamma_{A}P_{h}U_{ja}\nu_{a})[\overline{d}\Gamma_{A}(\alpha P_{L}
+\beta P_{R})u]^\dagger+h.c.\label{eqone} 
\end{equation}
and 

\begin{equation} \-$\it{$\mathcal{L}$}$^L=2\sqrt{2}G_{F}F^{hh'}_{Aijkm}
(\overline{l_{i}}\Gamma_{A}P_{h}U_{ja}\nu_{a}) (\overline
{l_{k}}\Gamma_{A}P_{h'}U_{mb}\nu_{b})^\dagger \label{eqtwo}, 
\end{equation}
where $U_{ia}$ is the matrix, unitary in the relativistic limit, that relates
flavor as shown below in Eq.(7).  Repeated indices are summed.
The coefficients K and  F, whose indices are i,j,k, ..., represent the 
coupling strengths for the different lepton flavor combinations, while the
indices a, b, ... label the mass eigenvalues.  The coefficients $\alpha$ and
$\beta$ allow for different strengths for L and R couplings to the quark
currents.
The Lorentz structure of the bilinear forms is labeled by A = S,V or T and 
$P_{h}$ denotes the  left- and right-helicity projections. 
The expression in Eq.(2) is a generalization of the generic muon-decay, four-
Fermi interaction to include lepton flavor violations of all types.  By a 
Fierz transformation one can show that $F_{T}^{LL}$ and $F_{T}^{RR}$ are both
identically zero.  Restricting application of Eq.(2) to muon decay and
(unobserved) massless neutrinos, one can show that it is not possible to test
lepton number conservation from the available observables \cite{LaLo}.

\medskip

We illustrate the notation by applying it to the standard model
(SM) effective, low-energy Lagrangian.   
The leptonic neutral-current term,

\begin{equation} \-$\it{$\mathcal{L}$}$_{SM}^{LNC}=\sqrt{2}G_{F}[\overline
{l_{i}}(2s_{W}^2\gamma_{\mu}P_{R}+\gamma_{\mu}(2s_{W}^2-1)P_{L})l_{i}]
(\overline{\nu_{j}}\gamma^{\mu}P_{L}\nu_{j}),\end{equation}
can be Fierz transformed into an equivalent charged current form

\begin{equation} \-$\it{$\mathcal{L}$}$_{SM}^{LNC}=\sqrt{2}G_{F}[[\overline
{l_{i}}(2s_{W}^2-1)\gamma_{\mu}P_{L}\nu_{j}](\overline{\nu_{j}}\gamma^{\mu}
P_{L}l_{i})-2[\overline{l_{i}}2s_{W}^2P_{L}\nu_{j}](\overline{\nu_{j}}P_{R}
l_{i})].\end{equation}
Next we add the SM leptonic charged-current effective
Lagrangian

\begin{equation} \-$\it{$\mathcal{L}$}$_{SM}^{LCC}=2\sqrt{2}G_{F}{[\overline
{l_{i}}\gamma_{\mu}P_{L}\nu_{i}](\overline{\nu_{j}}\gamma^{\mu}
P_{L}\l_{j})},\end{equation}
and the semileptonic effective  Lagrangian

\begin{equation} \-$\it{$\mathcal{L}$}$_{SM}^S=2\sqrt{2}G_{F}{[\overline
{u}\gamma_{\mu}P_{L}d](\overline{l_{j}}\gamma^{\mu}
P_{L}\nu_{j})}+h.c.\end{equation}
The F and K coefficients can now be read off from the SM effective low energy 
Lagrangian:
\begin{center}

\begin{tabular}{|l|l|l|l|r|} \hline
$K_{Vjj}^L$  & $F_{Viijj}^{LL}(i\neq j)$ & $F_{Viiii}^{LL}$ & $F_{Vijji}^{LL}
(i\neq j)$ & $F_{Sijij}^{LR}$\\ \hline
1 & 1 & ($s_{W}^2$+1)/2 & (2$s_{W}^2$-1)/2 & -2$s_{W}^2$\\ \hline
\end{tabular}
\end{center}.

\subsection{Lepton Flavor-Changing Transitions}

In the usual analysis of lepton flavor oscillations, the neutrinos are
treated as massless in the matrix element kinematics. 
Approximating the plane wave phase factors for the propagating
neutrino to leading order in the masses, one factors out the transition
amplitude to write\footnote{In the ultra-relativistic limit, one may take
t=L, the propagation length, in the following expressions.}

\begin{equation}\langle\nu_{j}(t)|\nu_{i}(0)\rangle=\sum_{a}\langle\nu_{j}|
\nu_{a}\rangle e^{-im_{a}^2 t/2E} \langle\nu_{a}|\nu_{i}\rangle=\sum_{a}
U_{ja}^*e^{-im_{a}^2 t/2E} U_{ia}.\end{equation}
However, this factorization is not valid when the
neutrino masses are taken into account \cite{KIM1}.  Though we work in the
ultra-relativistic limit, where the neutrino masses can be ignored in the
arguments of the matrix elements, we find the process dependence as discussed
in \cite{KIM1}to be a useful setting for the intermediate stages of our
development.  We generalize the SM weak-process initial states
to include new physics; for example,

\begin{equation} |\nu_{\mu}\rangle_{WP}\sim\sum_{a}|\nu_{a}\rangle
\langle \nu_{a},\mu^+|$\it{$\mathcal{L}$}$_{SM}^S|\pi^+\rangle\rightarrow 
\sum_{a}|\nu_{a}\rangle
\langle \nu_{a},\mu^+|$\it{$\mathcal{L}$}$^S|\pi^+\rangle, 
\end{equation}
with $L^S$ defined in Eq.(1).  Similarly, we create a
weak-process final state for the detector and use it with Eq.(8) to construct
the transition amplitude.  More generally, denoting initial and final states
by $|I^{s,d}\rangle$ and $|F^{s,d}\rangle$, we define

\begin{equation}$\it{$\mathcal{M}$}$_{a}^s=\langle F^{s} (\nu_{a})|$\it
{$\mathcal{L}$}$^S+$\it{$\mathcal{L}$}$^L|I^s\rangle,\end{equation}
and

\begin{equation}$\it{$\mathcal{M}$}$_{a}^d=\langle F^{d}|$\it{$\mathcal
{L}$}$^S+$\it{$\mathcal{L}$}$^L|I^d(\nu_{a})\rangle,\end{equation}
as the source and detector transition matrix elements involving a mass
eigenstate of $\nu_{a}$. We can write the full transition amplitude,
including direct new interactions, from
creation to detection of the neutrino as

\begin{equation}\langle \nu_{j}(t)|\nu_{i}(0)\rangle_{NI}=\sum_{a} 
\langle \nu_{j}|\nu_{a}\rangle e^{-im_{a}^2 t/2E}\langle \nu_{a}|
\nu_{i}\rangle_{NI}=$\it{$\mathcal{M}$}$_{a}^d e^{-im_{a}^2 t/2E}
$\it{$\mathcal{M}$}$_{a}^s,\end{equation}
where a sum over the label, a, of mass eigenstates is implicit here
and in what follows, and $NI$ indicates that new interactions are included.

\medskip

As an example, let us consider a SM process that is a background to electron
appearance experiments\footnote{These are experiments that look for signals
of neutrino flavor j interactions in a beam of neutrinos created with flavor
i $\neq$ j.}, namely $\pi^+\rightarrow e^+\nu_{e}$
followed by $ \nu_{e} N_{i}\rightarrow e^- N_{f}$, where $N_{i,f}$ designate
inital and final hadronic (nuclear) states.  We expand the SM
to include the possibility of neutrino oscillation for illustration.
The transition matrix elements for the source and detector processes are

\begin{equation}$\it{$\mathcal{M}$}$_{a}^s=2\surd 2 G_{F}K_{V11}^L\langle e^+
\nu_{e}|(\overline l_{1}\gamma_{\mu}P_{L}U_{1a}\nu_{a})^{\dagger}|0\rangle
\langle0|$\it{$\mathcal{O}$}$_{V}^{\mu}|\pi^+\rangle\end{equation}
and

\begin{equation}$\it{$\mathcal{M}$}$_{a}^d=2\surd 2 G_{F}K_{V11}^L\langle N_{f}|$\it{$\mathcal{O}$}$_{V}^{\mu\dagger}|N_{i}\rangle \langle e^-|\overline l_{1}
\gamma_{\mu}P_{L} U_{1a}\nu_{a}|\nu_{e}\rangle,\end{equation}
where \begin{math}\it{\mathcal{O}}_{V}^{\mu}\end{math} designates the
quark current operator
$\overline d\gamma^{\mu}P_{L}u$ appropriate to the SM in Eq.(1). In the
following, we will continue to use $\it{\mathcal{O}}$ to designate
the hadronic current operator.  In the
ultra-relativistic limit, the neutrino masses are set equal to zero in the
spinors and only the leading phase dependence on masses is kept. The
transition amplitude squared can then be factorized into a SM product of matrix
elements squared times the oscillation probability, involving a sum over
mass eigenstates:

\begin{equation}|$\it{$\mathcal{M}$}$_{a}^d e^{-i m_{a}^2 t/2E}
$\it{$\mathcal{M}$}$_{a}^s|^2 = (2\surd 2 G_{F})^4|(\overline e\gamma_{\mu}
P_{L}\nu)^*(\overline e\gamma_{\lambda}P_{L}\nu)\langle 0|$\it{$\mathcal
{O}$}$_{V}^{\mu}|u\overline d\rangle\langle N_{f}|$\it{$\mathcal
{O}$}$_{V}^{\lambda\dagger}|N_{i}\rangle|^2 |U_{1a}^*e^{-im_{a}^2 t/2E}
U_{1a}|^2,\end{equation}
where we have used the fact that $K_{V11}=1$ in the SM. 
Schematically we can write the above as

\begin{equation}|$\it{$\mathcal{M}$}$_{a}^d e^{-i m_{a}^2 t/2E}
$\it{$\mathcal{M}$}$_{a}^s|^2 \sim \Phi_{e}\sigma(\nu_{e} N_{i}\rightarrow 
e^{-} N_{f})P_{e\rightarrow e}.\end{equation}
In Eq.(15) $\Phi_{e}$ designates the $\nu_{e}$ flux, $\sigma(\nu_{e} N_{i}
\rightarrow e^- N_{f})$ the SM electron neutrino charged current cross section
and $P_{e\rightarrow e}$=$|U_{1a}^* e^{-im_{a}^2 t/2E}U_{1a}|^2$ designates
the probability that an electron neutrino produced at
the source appears as an electron neutrino at the target.  Our next task is to
achieve an equally transparent factorization that includes direct violation of
lepton flavor.

\subsection{Lepton Flavor-Changing Probability Factor}

To identify a flavor-changing factor that includes direct flavor violation
effects and plays the same role as the oscillation
probability factor $P_{e\rightarrow e}$ in Eq.(15), we use the amplitudes
defined in Eqs.(9), (10) and (11) and follow the pattern of our preceding
discussion of Eqs.(12) - (15).  Our new ``probability'' factor should reduce
to the oscillation probability when the direct flavor-violating couplings are
turned off, of course. For this purpose, we continue to use the pion
decay source and nucleon/nucleus detector example,
but now we will extract a lepton appearance probability
factor from combined oscillations and direct interactions. We adopt an
illustrative model
with a V-A current with $K_{V2l}^L\neq 0$, with $l = \tau$ or
e but with all other non-SM coefficients $K_{Aij}^h$=0. This model produces a
$\pi^+\rightarrow\mu^+ +\nu_{l}$ transition at the source but leaves only the
SM process $\nu_{l}+N_{i}\rightarrow l^- +N_{f}$ active at the target. We
shall refer to this as a ``source-only'' situation. The electron production
rate at the detector is proportional to

\begin{equation}|$\it{$\mathcal{M}$}$_{a}^d e^{-i m_{a}^2 t/2E}
$\it{$\mathcal{M}$}$_{a}^s|^2=|\langle l^- N_{f}|$\it{$\mathcal{L}$}$^S|
\nu_{a}N_{i}\rangle e^{-im_{a}^2 t/2E}\langle\mu^+ \nu_{a}|$\it{$\mathcal
{L}$}$^S|\pi^+\rangle|^2,\end{equation}
where the right-hand side, after spelling out the terms from Eq.(1) and taking 
the ultra-relativistic limit, reads

\begin{equation}
(2\surd 2 G_{F})^4|\overline l\gamma_{\lambda}P_{L}\nu\langle N_{f}|
$\it{$\mathcal{O}$}$_{V}^{\lambda}|N_{i}\rangle|^2 |\langle 0|
$\it{$\mathcal{O}$}$_{V}^{\sigma}|\pi^+\rangle
\overline\nu\gamma_{\sigma}P_{L}\mu|^2
|(U_{2a}+K_{V2l}^L U_{la})e^{-im_{a}^2 t/2E}
U_{la}^*|^2,\end{equation}
where $K_{V11} = 1$ has been used.
The $\nu_{l}$ appearance probability factor, including the direct flavor-
violation coefficient $K_{V2l}^L$, is given by the final expression 
within absolute magnitude signs in Eq.(17).  The first term in this expression
is the usual oscillation amplitude for transition from $\nu_{\mu}$ to 
$\nu_{l}$ between source and detector.  

To make the notation more compact in the following discussion, we define

\begin{eqnarray}(m_{b}^2-m_{a}^2) t/2E&\equiv&2x\nonumber \\ K_{V2l}^L&\equiv&
tan\psi e^{2i\phi},\nonumber\end{eqnarray}
where the choices of a and b in the definition of x and the values of $\psi$
and $\phi$ depend upon the situation, 
and we will illustrate the two-flavor mixing and the three-flavor mixing
with dominance by one mass scale\cite{cf96}. 
We have chosen the parametrization of
$K_{V2l}^L$ by $tan\psi$ to emphasize that the flavor-violating amplitude
could,
in principle, be larger than the flavor-conserving one.  In practice it
is constrained by experiment to be small compared to the standard model
amplitude, but the parametrization reminds us that the direct flavor-
violating amplitude cannot be represented as a unitary rotation to a ``source
basis'' except in the small $tan\psi$ limit with $\phi$ negligible.  In 
addition, the special circumstance must occur
that the effective flavor-violating four-Fermi interactions
have a V-A structure to match the SM structure.  Nonetheless, we will refer
to the transition factor modulus squared as a probability or probability
factor in what follows.  As we show below, the survey of direct effects
on the standard two- and three-family mixing plots with $\Delta m^2$ vs.
$sin^2(2\theta)$ contours or $tan^2 \theta_{13}$ vs. $tan^2\theta_{23}$
for fixed $\delta m^2$, for example, can be straightforwardly extended to 
include direct lepton number violation. 

 Isolating the probability factor in Eq.(17), we
write 

\begin{equation}P_{\mu\rightarrow l}=|U_{2a}e^{-im_{a}^2 t/2E}U_{la}^*
 +tan\psi e^{2i\phi} U_{lb}e^{-im_{b}^2 t/2E}U_{lb}^*|^2,\end{equation}
no sum on $l$, which labels e or $\tau$.
With $\phi=0$, in ``mock'' unitary form,\footnote{For present purposes,
we assume $\phi$ is
constrained to be small by experimental limits on CP violation.} we have

\begin{equation}P_{\mu\rightarrow l}=|cos\psi U_{2a}e^{-im_{a}^2 t/2E}U_{la}^*
+sin\psi U_{lb}e^{-im_{b}^2 t/2E}U_{lb}^*|^2 /(cos\psi)^2,
\end{equation}
and, approximating the overall $(cos\psi)^{-2}$ factor by 1 to order $\psi^2$, 
Eq.(19) can also be recast as
 
\begin{equation}P_{\mu\rightarrow l}\simeq |V_{22}U_{2a}e^{-im_{a}^2 t/2E}
U_{la}^*+V_{2l}U_{lb}e^{-im_{b}^2 t/2E}U_{lb}^*|^2.\end{equation}
We have identified $V_{22}=cos\psi$ and $V_{2l}$=$sin\psi$ in Eq.(20), which
suggests the interpretation that the $\mu$ from $\pi$ decay
is accompanied by a $\nu_{\mu}$ with amplitude $V_{22}$ and followed by
propagation and
an oscillation to $\nu_{e}$ with amplitude $U_{2a}e^{-im_{a}^2 t/2E}U_{1a}^*$
or is accompanied by a $\nu_{l}$ $l\neq \mu$ with amplitude $V_{2l}$ and then
followed by $\nu_{l}$
propagation with amplitude $U_{lb}e^{-im_{b}^2 t/2E}U_{lb}^*$ to remain
$\nu_{l}$. 
This picture, though only approximately valid and only then 
in the circumstances described above, is useful for seeing how the
direct lepton flavor violations can be worked in with the oscillation between 
flavors in a reasonably seamless fashion.

\section{Constraints on the Strengths of Direct Lepton-Flavor Violations}

The identities of electron neutrino, muon neutrino and tau neutrino
are preserved to a good accuracy in decays and collisions.  
Within neutrino data alone, the constraints on the sizes
of lepton-flavor violations, though tight, often leave room for violations at
the 1/2\% level in cross-sections and rates.  In the realm of purely charged
lepton processes, on the other hand, the high degree of experimental control
has lead to fantastically small limits on some of the ratios of lepton-number-
violating to lepton-number-preserving interaction strengths.  In this
section we briefly survey the constraints that affect our analysis the most
directly.  The considerations presented here govern our choice of parameter
values in the following sections as we illustrate some of the possible effects
that direct interactions produce in concert with oscillations.  

Weak SU(2) invariance of any new physics interaction Lagrangian that 
produces lepton number violation will generally relate the purely leptonic
processes to those involving neutrinos\cite{BG}.  The severe experimental
limits on the kinematically allowed leptonic processes then translate into
limits on processes involving leptons and neutrinos, up to group theory factors
and SU(2)-violating mass-splittings among members of boson multiplets that
mediate the lepton number violations.  The purely leptonic processes whose
experimental limits
impose the strictest bounds on lepton number violation in muon-source 
experiments are muonium to antimuonium conversion\cite{mumu},
$\mu\rightarrow eee$,
$\tau\rightarrow\mu ee$ and $\tau\rightarrow\mu\mu e$\cite{PDG} . 
There are similar constraints\cite{PDG} on the strengths of semileptonic
processes violating lepton number that follow from $\tau\rightarrow \pi+e$, 
$\tau\rightarrow \pi+\mu$ and from $\mu$ conversion to e on Ti nuclei
\cite{kkl}.  The constraints are summarized in Table 1.
The ``experimental constraints'' shown in Table 1 are actually those that
apply to the 
charged-lepton processes listed.  The relationship to the bounds on the
lepton-number-violating coefficients for the neutrino processes is somewhat
indirect, since group theory factors, ratios of masses of virtual bosons
mediating the processes and effects due to differences between, say, S and
V structure must be included.  Allowing a possible factor of two from 
Clebsch-Gordon coefficients and a generous factor of two in the ratio of
masses of exchange bosons within the same SU(2) multiplet, the coefficients
shown in Table 1 are bounded by roughly eight times the purely charged-
lepton process limits listed there\cite{BG}. This value is given in the 
column labeled ``model independent constraint''.
The precise value of the factor for a given process is model dependent.
In Table 1 and in the rest of the paper, a superscript L is to be 
understood if none is shown explicitely.
\begin{table}[h]
\begin{tabular}{|l|l|l|l|r|}                                         
\emph{coefficient} &\emph{process} &\emph{experimental constraint} &\emph
{model independent constraint}\\ \hline
  $F_{V2111}$      & $\mu\rightarrow eee$       & $1.0\times10^{-6}$ 
& $8.0\times10^{-6}$ \\   
  $F_{V2112}$      & muonium-antimuonium        & $3.0\times10^{-3}$
& $2.4\times10^{-2}$  \\   
  $F_{V2113}$      & $\tau\rightarrow \mu^+ ee$ & $2.9\times10^{-3}$
& $2.3\times10^{-2}$  \\
  $F_{V2213}$      & $\tau\rightarrow e^-\mu\mu$& $3.3\times10^{-3}$
& $2.6\times10^{-2}$  \\
  $F_{V2311}$      & $\tau\rightarrow \mu^- ee$ & $3.2\times10^{-3}$ 
& $2.5\times10^{-2}$ \\
  $F_{V2312}$      & $\tau\rightarrow e^+\mu\mu$& $2.9\times10^{-3}$
& $2.3\times10^{-2}$  \\      
  $K_{V21}$        & $\mu-e$ conversion         & $1.8\times10^{-7}$ 
& $1.5\times10^{-6}$ \\
  $K_{V31}$        & $\tau\rightarrow e\pi^0$   & $8.2\times10^{-3}$ 
& $6.6\times10^{-2}$ \\
  $K_{V32}$        & $\tau\rightarrow \mu\pi^0$ & $8.5\times10^{-3}$
& $6.8\times10^{-2}$  \\ 
\end{tabular}
\caption{Limits on charged lepton processes and the corresponding neutrino
process coefficients F and K.  The model independent constraints on the F and K
coefficients are taken to be roughly eight times the charged lepton process
limits.  For K constraints we take h=L, $\alpha$=1 and $\beta$=0 in Eq.(1).}
\end{table}  
  
\section{Impact on Analysis of Appearance Experiments}

The very term ``neutrino oscillation'' implies path-length-dependent
variation of the probability that a given flavor of neutrino appears in
the beam.  Moreover, neutrino oscillation and neutrino mass are so tightly
linked that evidence for the former is considered tantamount to proof of the
latter - certainly in the case of vacuum oscillations. 
Conversely, the absence of oscillations in a neutrino flavor-violating
effect is tantamount to elimination of neutrino mass as
an explanation of its origin. This is not necessarily so when flavor
violation is expanded to include direct interactions.  This point is among
a number that we make in the present section.  The examples chosen are all
consistent with the bounds described in the previous section and summarized
in Table 1.

At the beginning of this section we take the neutrino source to be $\pi$-
decay.  We explore the interplay in the $\nu_{\mu}\leftrightarrow \nu_{\tau}$
case between the mass-induced, oscillating amplitude and the directly-induced,
non-oscillating amplitude.  While it is true that small masses generally
lead to vacuum oscillations, it is not strictly true
that the absence of oscillations proves that the neutrinos are massless. In
a sense this is a complement to the well-known result that massless neutrinos
can oscillate as they pass through matter.  These effects are shown in Figs.
1. and 2.  In Fig. 3., we show the result of including the flavor-violation
parameter ``axis'', $\psi$, in the analysis of the probability bounds in the
sensitive NOMAD experiment.  The result is dramatic.  Then we make the point
that the bound on the oscillation mixing angle for fixed $\Delta m^2$ depends
in general on the value of the direct flavor violation parameter $\psi$.  
This is shown in Fig. 4, which includes the possibility that the same direct
flavor violation occurs at both the source and the detector.  This discussion
is followed by the expansion of the analysis to the three flavor situation.
Figures 5. and 6. show how different $tan^2\theta_{13}$ - $tan^2\theta_{23}$
boundaries appear as different fixed $\psi$ and $\Delta m^2$ ``slices''
of the parameter space are taken.  Figure 5. is appropriate to the upcoming
MiniBooNe experiment, while Fig. 6. applies to the reported NOMAD probability
bound.

In the last part of this section, we show the power of the clean $\nu_{\mu}$
and $\nu_{e}$ beams from proposed $\mu^+/\mu^-$ storage rings to make sensitive
tests for direct flavor violation in the purely leptonic sector. Of particular
note is the prospect of advancing another order of magnitude into the parameter
space of direct $\nu_{e}\leftrightarrow\nu_{\mu}$ flavor violations.

\subsection{ $\pi$-decay as the Neutrino Source}    

Restricting ourselves at first to mixing of two mass eigenstates, we write 
Eq.(19) for the probability of lepton appearance in the pion decay
as the source in an ``all angles'' form and defining 
$F^L_{V2l}=tan\psi e^{2i\phi}$ 

\begin{equation}P_{\mu\rightarrow l}=
|cos\psi(-cos\theta sin\theta e^{ix}+sin\theta cos\theta 
e^{-ix})+sin\psi e^{2i\phi} (cos^2 \theta e^{ix}+sin^2 \theta e^{-ix})|^2
cos^{-2}\psi.\end{equation}
As defined above, $x=(m_{2}^2-m_{1}^2)t/4E$ in applicaton to Eq.(21),
and the two-flavor mixing matrix is written as

\begin{equation}U=\left( \begin{array}{cc} \cos\theta & \sin\theta \\
-\sin\theta & \cos\theta \end{array}\right). \end{equation}
Rearranging terms and consolidating them, we arrive at the rather transparent
form of Eq.(21)

\begin{equation}P_{\mu\rightarrow l}=tan^2\psi+\frac{sin2\theta sin
2(\theta-\psi)sin^2 x}{cos^2\psi}+4 tan\psi sin 2\theta sin\phi sinx(
cos2\theta sin\phi sinx-cos\phi cosx).\end{equation} 
Equation (23) has the  obvious and expected feature that if $\psi$=0, we
have $P_{\mu\rightarrow l}=sin^2 2\theta sin^2 x$, the usual two-flavor, pure
mass-mixing, oscillation formula in terms of the mixing angle $\theta$ and
the factor $x=\Delta m^2 L/4E$.  Equally obvious and expected is the
relationship 
$P_{\mu\rightarrow l}=tan^2\psi$ that holds when $\theta=0$ or $x=0$.
\emph
{What is not expected is that, 
when $sin\phi =0$,
$P_{\mu\rightarrow l}=tan^2\psi$, independent of x, 
when $\theta-\psi=\pi n/2$}, where n is an integer. Looking back at Eq.(21),
we see that when $\theta-\psi=\pi n/2$ and when $\phi$=0 the coefficient of
$e^{ix}$ from the $\nu_{\mu}\rightarrow\nu_{l}$ oscillation cancels against
its coefficient from the $\nu_{l}\rightarrow\nu_{l}$ term in the direct
flavor violation amplitude.  The remaining overall phase from the $e^{-ix}$
factor disappears in the modulus squared and one is left with simply
$P_{\mu\rightarrow l}=tan^2\psi=tan^2\theta$.  This somewhat surprising result
in the case where there is direct flavor violation in the DIF source dramatizes
the implications of Eq.(23) for interpreting signals for oscillation,
or lack thereof, in variable baseline experiments. There are counterparts to
this source effect in the muon decay case \cite{ladar}\cite{krdar}
as well as in
the cases where direct flavor violation occurs only at the detector or in
both the detector and the source.  We will comment further on these 
situations and on three-flavor mixing below.
 
The condition for exact cancellation of the L/E dependence is unlikely, of
course, but the interplay between the $\psi$ and $\theta$ dependence
is generic, and it affects, possibly radically if mixing angles are small,
the interpretation of signals that show variation with path length. We 
illustrate the L/E-dependent effects caused by the interference between
the pure oscillation term and the direct flavor violation in Figs. 1 and
2. 
\begin{figure}[t,b]\hbox{\hspace{6em}
\hbox{\psfig{file=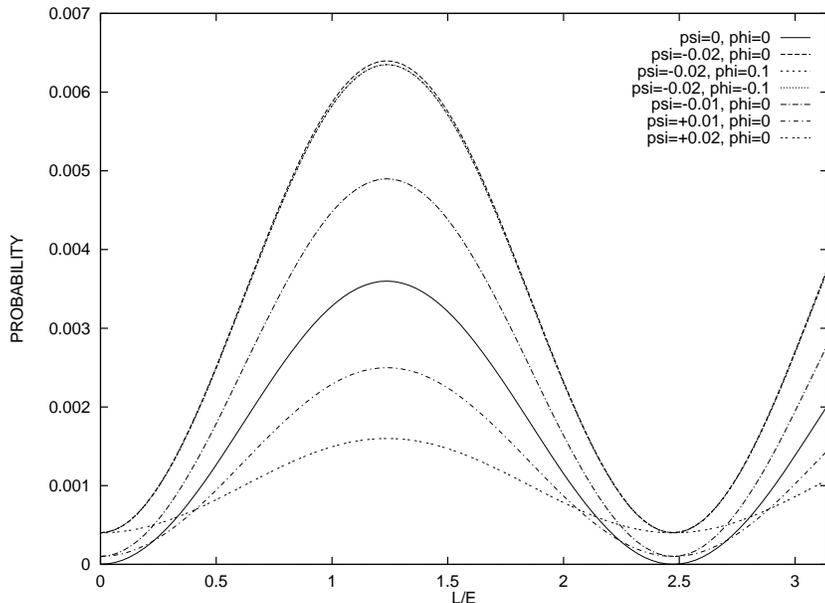,height=8 cm,angle=-90}}}
\caption{Dependence of $P_{\mu\rightarrow l}$ on $x$
for $\theta=0.03$ and several different values of $\psi$
and with $\Delta m^2 = 1ev^2$. A Gaussian smearing model has 
been adopted with $\sigma=0.016$, appropriate to the NOMAD experiment,
which we will
use as an illustrative example.  The top three curves, which are barely
distinguishable from one another, show the very weak
dependence on the CP-violating phase $\phi$}\end{figure}
Figure 1 illustrates the variation of $P_{\mu\rightarrow l}$ with L/E for
fixed $\theta=0.03$, or $sin^2 2\theta=0.0036$, for various values of $\psi$. 
Figure 2 shows the variation for $\psi=-0.02$ and various
values of $\theta$.  Several combinations of $\theta$ and $\psi$
can lead to a given
curve, which suggests that it would not be straightforward to
disentangle the oscillation parameters from the comparison of the
$x$-dependence of the probability with data if small direct effects were 
included in the analysis.
\begin{figure}[t,b]\hbox{\hspace{6em}\hbox{\psfig{file=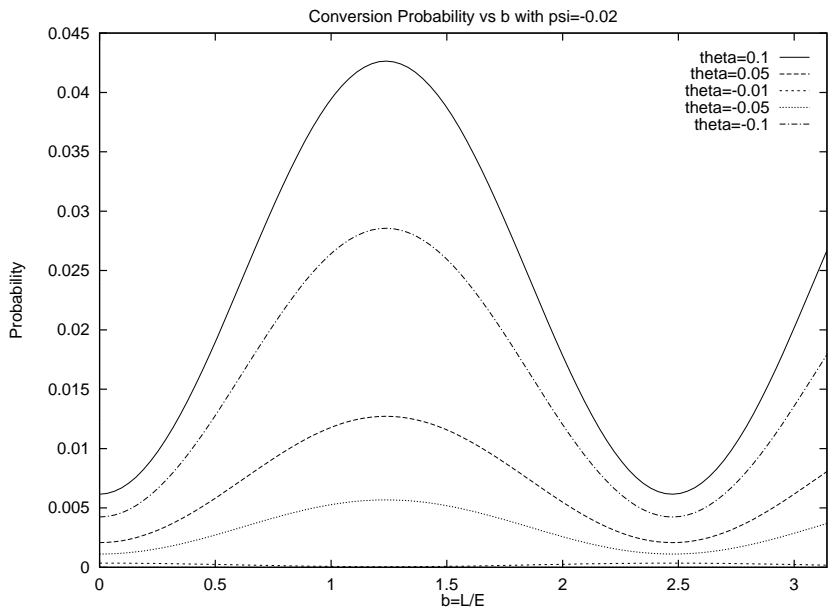,height=
8 cm,angle=-90}}}\caption{Dependence of $P_{\mu\rightarrow l}$ on $x$
for $\psi=-0.02$ and several different values of $\theta$ and
with $\Delta m^2 = 1ev^2$.}\end{figure}

For completeness, we also show in Fig. 1 the (weak) influence of $\phi$
on the two-flavor oscillation amplitude.  Comparing the top three curves,
we see that including the $\phi$-dependent
effects for reasonable $\phi$ values leaves the picture essentially unchanged,
as one expects when a small CP-violating phase rests upon a small
flavor-violating amplitude.  A CP-violating phase of order one is needed if
it is to make an observable impact. We reserve the analysis of effects from
large CP-violating phases\cite{kanecp} for a future study.

We see that the behavior of the probability is governed essentially by the
second term in Eq. (23).   This term
makes it clear that even small $\psi$ values can have a large effect if the 
mixing angle $\theta$ is of the same order of magnitude as $\psi$. 

Are there circumstances where the sizes of these direct effects could be
big enough to be observable?  A glance at Table 1 shows that
$K_{V21}$ (=$tan\psi$) is
constrained to be far too small, even with generous allowance for SU(2)
breaking effects, to modify the pure oscillation picture of $\nu_{\mu}$
from $\pi$ decay oscillating to $\nu_{e}$.  The $K_{V32}$ bound is much
looser, however, and interesting effects could occur in
$\nu_{\mu}\rightarrow \nu_{\tau}$
two-flavor mixing with $tan\psi\leq 0.02$ if $\theta_{\mu\tau}$ were of the
order of 0.1 or less. A large mixing angle is required between $\nu_{\mu}$ and
another species by the two-flavor fit to the atmospheric neutrino anomaly.
Thus mixing with a sterile neutrino is required, in order to have 
a direct interaction
effect that shows up in a (weakly) mixed $\nu_{\mu}\leftrightarrow
\nu_{\tau}$ sector and significantly modifies the $\nu_{\tau}$
appearance signal from a pion decay source of $\nu_{\mu}$. 
For small appearance probabilities, the change in signal depends rather
sensitively on the value of the direct interaction strength.  We give some
detail in an example in the $\mu$ decay case in subsection B below.

\subsubsection{Impact on Boundaries in $\mu-\tau$ Mixing Space}

Failure to detect $\nu_{\tau}$'s in an appearance search allows one to set
confidence level curves in $\mu$-$\tau$ mixing space, and this gives another
slant on the application of our formalism.  The recent NOMAD results\cite{NOM}
give the smallest probability, and tightest large $\Delta m^2$ limits on the
allowed region of parameter space for two-flavor, $\nu_{\mu}\leftrightarrow
\nu_{\tau}$ mixing.  To illustrate the impact of small direct interaction
effects on the NOMAD bound, we approximate their boundary curve, which 
corresponds to a P=0.0006 appearance probability, by the simple Gaussian
smearing model\cite{PDG} with parameters fit to reproduce the main features
of the NOMAD boundary in $sin^2\theta$ vs. $\Delta m^2$ plane for mixing
of $\nu_{\mu}$ and $\nu_{\tau}$.
Keeping these parameters fixed, we replot the contour in the
$sin^2 2\theta$-$\Delta m^2$ plane for several small values of $\psi$. 
As before the small $\phi$ effects are not interesting for our present point,
and we set $\phi$=0.  The result is shown in Fig. 3. 

\begin{figure}[t,b]\hbox{\hspace{6em}\hbox{\psfig{file=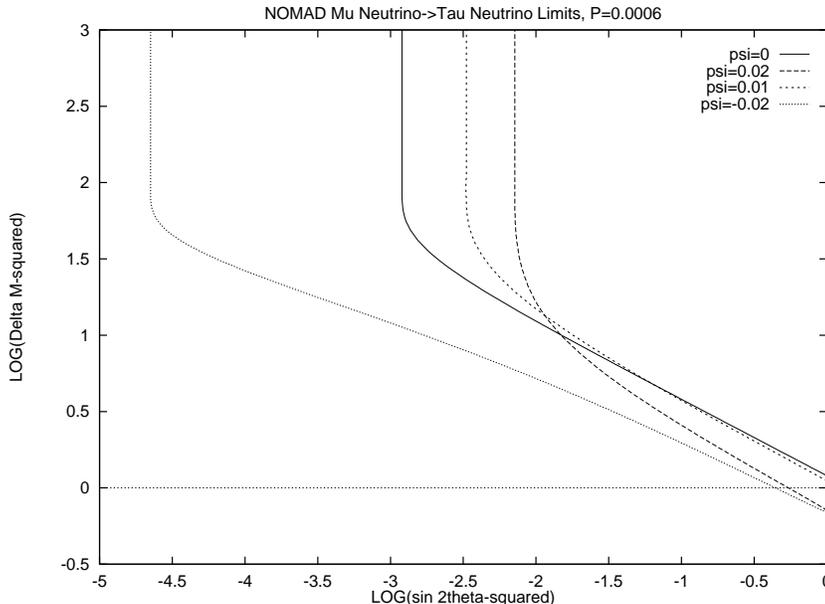,height=
8 cm,angle=-90}}}\caption{Model of the NOMAD $90\%$ C.L. limit on
$\nu_{\mu}\rightarrow\nu_{\tau}$ oscillations for several values of $\psi$
with $\phi$=0.  The curves correspond to P=0.0006.}\end{figure}
The boundaries of the null search results are significantly revised  even when
small direct effects, consistent with
the bounds from other experiments, are introduced.  NOMAD's high sensitivity to
small appearance probability in the large $\Delta m^2$ region is the reason
that the inclusion of small direct amplitudes has such a pronounced effect.

A few comments on the shape of the different curves is in order.  NOMAD has
a small average L/E value, so it is sensitive only to large $\Delta m^2$.
Their L/E distribution is broad, so the smearing almost completely damps 
the oscillations in the boundary curve at large $\Delta m^2$. The Gaussian
model is even more extreme in this respect than the actual fit to the data.
Because of the factor $sin 2(\theta-\psi)$ in the expression for 
$P_{\mu \rightarrow \tau}$, for $0\leq\theta \leq \pi/2$ the positive
$\psi$ values reduce the amplitude and the $sin^2 x$ factor has to compensate.
This eventually drives the curve to higher $\Delta m^2$ for a given $\theta$
value.  The opposite behavior occurs when $\psi$ is negative.  The role of
the term $tan^2\psi$ is most apparant at large values of $\theta$, where the
$\psi$ in the argument of sine is inconsequential, but the addition of 
$tan^2\psi$ to the probability allows a fit with somewhat smaller $\Delta m^2$
at $\theta=\pi/2$.  The crossing of positive $\psi$ curves with the $\psi$
=0 curve is forced by the large $\Delta m^2$ and small $\Delta m^2$ behaviors
just described.
 
\subsubsection{Direct Flavor-Change Bounds in $\psi-\theta$ Space}

Proposed high-sensitivity experiments to probe smaller $\Delta m^2$ and
$sin^2 2\theta$ regions can also place discovery limits and upper bounds on
direct, flavor-violating interactions involving neutrinos. 
For example, Fermilab
\cite{MBNE} and CERN \cite{Ludov} proposals aim to push $P_{\mu\rightarrow e}$
bounds down to $10^{-4}$ at $90\%$ C.L., while the Fermilab-Soudan
experiment, MINOS \cite{MIN}, is shooting for a bound of $10^{-2}$ on
$P_{\mu\rightarrow\tau}$.  Similar sensitivity is proposed in muon collider
sources of pure $\nu_{e}$ and $\nu_{\mu}$ beams. 

Let us consider the situation where $\nu_{\mu}$'s originate from $\pi$ decays,
which will be the case in the MiniBooNE experiment\cite{MBNE}. 
If the new flavor physics is 
only at the source, Eq.(23) applies.  If one has the same new physics 
amplitude at the source and detector, the corresponding expression is

\begin{equation}P_{\mu\rightarrow l}=4tan^2\psi cos^2 x + sin^2 2\theta
 sin^2x - 2tan\psi sin2\theta sin2\phi sin2x.\end{equation}
Assuming that the CP-violating phase, $\phi$, is small, we can readily 
illustrate the influence of the x and $\theta$ values on the limits on $\psi$
imposed by a given $90\%$ C.L. bound on $P_{\mu\rightarrow l}$.  In Fig. 4
we show the boundaries in $\psi-\theta$ space for a fixed
value of x=1.27(L/E)$\Delta m^2$=1 when $P_{\mu\rightarrow l}\leq 10^{-4}$ is
imposed.  Fig 4 gives the source-only case and the
source-plus-detector case boundaries.  The specifications of MiniBooNE are
$L\simeq$500m and typically 0.5$\leq$E$\leq$1.0 GeV \cite{MBNE}, so
x$\simeq\Delta m^2$ for 
purposes of translating the graphs to MiniBooNE's capabilities.
 
\begin{figure}[t,b]\hbox{\hspace{6em}\hbox{\psfig{file=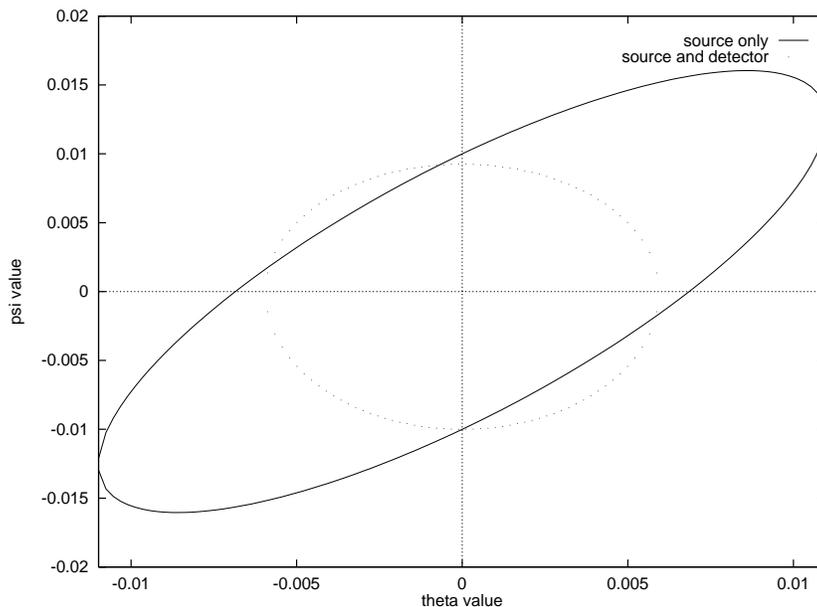,height=
8 cm,angle=-90}}}\caption{Contours of $90\%$C.L. upper bounds in
$\psi-\theta$ space for x=1 and $P_{\mu\rightarrow e}$
=$10^{-4}$. The cases where the new physics is at the source only and where it
is at both the source and detector with equal amplitudes are shown.}
\end{figure}

%\begin{figure}[t,b]\hbox{\hspace{6em}\hbox{\psfig{file=dnfig3c.ps,height=
%8 cm,angle=-90}}}\caption{Contours of $90\%$C.L. upper bounds in
%$\psi-\theta$ space at several x values.  In this case the new physics
%is the same both at the source and at the detector.}\end{figure}
In Fig. 4 the areas within the ``egg-shaped'' regions are allowed at the
$90\%$C.L., and the maximum allowed direct flavor violation parameter is 
evidently a function of both $\theta$ and x.  It is generally assumed that the
best bound on the flavor-violating amplitudes is obtained when the oscillations
are not present, but the ``tilted egg'' in Fig. 4 shows that this is not
necessarily the case. 
The interference between the oscillation and direct amplitudes in the
source-only situation makes the biggest direct flavor violation effect
occur at non-zero $\theta$.  When the same flavor violation is assumed to apply
at the source and the detector, the symmetry of the set-up ensures that the
maximum allowed value of $\psi$ occurs at $\theta$=0,
which is the usual expectation.

With the $\Delta m^2$ value chosen, the bound $|\psi|\leq$1.6$\times 10^{-2}$
results, which is an order of magnitude stronger than the current direct 
bound from neutrino processes, but still much weaker than the bound that can
be inferred from the limits on $\mu\rightarrow e$ transitions in heavy nuclei
as listed in Table I.

\subsubsection{The Three-Flavor System}

Distinct new possibilities are created when e, $\mu$ and $\tau$-flavors are all
included in the 
picture with mixing plus direct flavor violation.  Here we re-examine the
$\pi$-source, e and $\tau$ appearance possibilities with three flavors, using
the ``one mass-scale-dominance'' model \cite{cf96}.

Since the experimental constraint on $K_{V21}$ makes it irrelevant for our
purposes (see Table 1), we do not include it in the expressions below.  With
flavor violation at the source, and restricting ourselves to V-A structure, 
we need only consider the $K_{V22}$ and $K_{V23}$ coefficients.  To the first
order in flavor violation, we have

\begin{equation} P_{\mu\rightarrow e}=sin^2 2\theta_{13} sin^2(\theta_{23}+
\psi_{\mu})sin^2 x cos^{-2} \psi_{\mu}\end{equation}
and
\begin{equation} P_{\mu\rightarrow\tau}=tan^2 \psi_{\mu} + 4sin^2 x 
cos^2 \theta_{13} cos\theta_{23} cos^{-2} \psi_{\mu} sin(\theta_{23}+
\psi_{\mu})(-sin\psi_{\mu}+cos^2 \theta_{13}cos\theta_{23}sin(\theta_{23}+
\psi_{\mu})),\end{equation}
where $K_{23}\equiv tan\psi_{\mu}$ and the mixing-angle convention is that
of \cite{PDG}; namely $U_{13}$=$sin\theta_{13} e^{-\delta_{13}}$, $U_{23}$=
$sin\theta_{23} cos\theta_{13}$ and $U_{33}$=$cos\theta_{13} cos\theta_{23}$.
We have assumed $\delta_{13}$=0 and real $K_{Vij}$ values.  $P_{\mu\rightarrow
e}$ ($P_{\mu\rightarrow\tau}$) includes the amplitude that $\nu_{\mu}$ is 
produced at the source and oscillates to $\nu_{e}$ ($\nu_{\tau}$) plus the
amplitude that $\nu_{\tau}$ is produced at the source and oscillates to
$\nu_{e}$ (remains $\nu_{\tau}$).  Note that, as in the two-flavor case, the
special condition $\theta$+$\psi$=n$\pi$ kills the oscillation amplitudes,
and in the $\mu\rightarrow\tau$ case there is a second condition when this
can happen.  The lack of symmetry between $P_{\mu\rightarrow e}$ and
 $P_{\mu\rightarrow \tau}$ results from our neglect of the $K_{21}$ 
coefficient in the amplitudes.  Figure 5  shows the effect of choosing
different x values (i.e. different $\Delta m^2$ values for fixed L/E)
and non-zero $\psi_{\mu}$ values on the 
$P_{\mu\rightarrow e}$=$10^{-4}$ boundary in the
$tan^2 \theta_{13}$ vs.$tan^2 \theta_{23}$ plane, which is appropriate
for the MiniBooNE\cite{MBNE} parameters.

\begin{figure}[t,b]\hbox{\hspace{6em}\hbox{\psfig{file=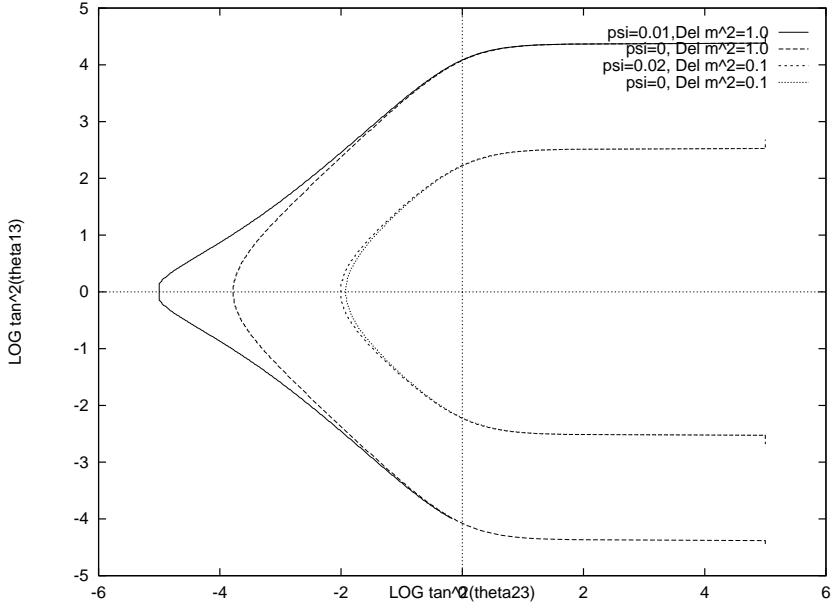,height=
8 cm,angle=-90}}}\caption{Contours of $90\%$C.L. upper bounds in
$tan^2 \theta_{13}$ vs. $tan^2 \theta_{23}$ space for several
sets of $\psi_{\mu}$, $\Delta m^2$ values and with fixed value 
$P_{\mu\rightarrow e}$=$10^{-4}$. The new physics effects are taken
to be at the source only.}
\end{figure}
For $\Delta m^2$=1$eV^2$ the change in the small $tan^2\theta_{23}$ region as
$\psi_{\mu}$ is
turned on is noticeable already at $\psi_{mu}$=0.01.  The distortion grows
rapidly with $\psi$, and the $LOGtan^2\theta_{23}$ minimum value
moves well below $10^{-6}$  when
$\psi_{\mu}$=0.02 and $\Delta m^2$=1. The symmetry of the graph about the
$tan^2\theta_{13}$=1 line is a consequence of the equivalence of the $\theta$
values below and above $\pi$/4 caused by the $sin^2 2\theta$ factor.

The influence of flavor violations in $\tau$-flavor processes is not so
tightly constrained experimentally as for strictly e-plus-$\mu$,
and we look at this situation next.  We contrast the bounds on 
$\tau$-appearance probability in the cases that the flavor violation is at
the source alone and the source and detector both.  In Fig. 6 we again look
at the $tan^2\theta_{13}$ and $tan^2\theta_{23}$ plane boundaries for
$P_{\mu\rightarrow\tau}$=$6\times10^{-4}$, as in the NOMAD experimental
boundary, discussed in IV.A.1 above in the two-flavor model. Plots with
$\Delta m^2$=9,6 and 3$eV^2$ are shown, with oscillation only in the
9 and 6 $eV^2$ cases but with $\psi_{\mu}$=0.02,0.0 and -0.02 for
the source-only case at $\Delta m^2$=3$eV^2$ and, for comparison, $\psi_{\mu}$
=0.01 and -0.01 in the source-plus-detector case at $\Delta m^2$=3$eV^2$.   

\begin{figure}[t,b]\hbox{\hspace{6em}\hbox{\psfig{file=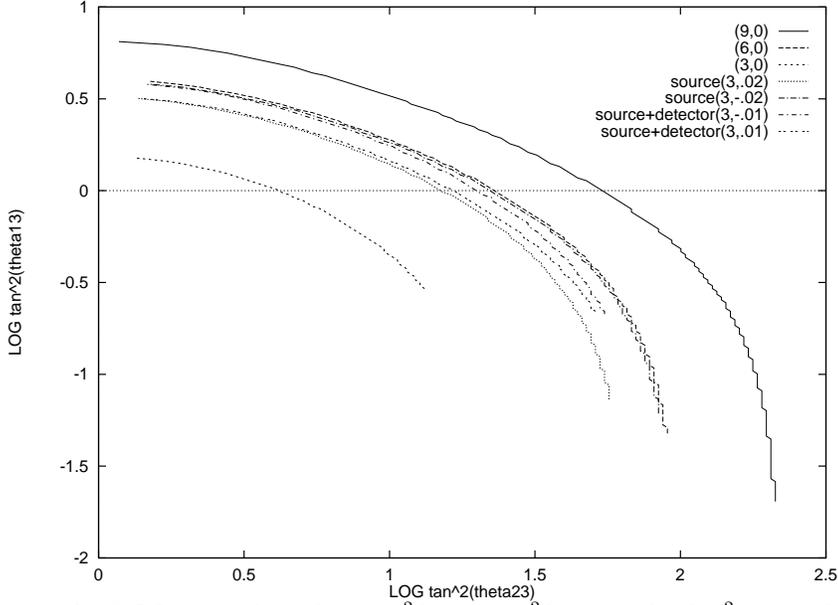,height=
8 cm,angle=-90}}}\caption{Contours of $90\%$C.L. upper bounds in
$tan^2 \theta_{13}$ vs. $tan^2 \theta_{23}$ space for $\Delta m^2$=9,6 and
3$eV^2$. The curves are designated as source or source and detector according
to whether flavor violation is included at the source only or both the
source and the
detector. In parethesis the first number is the $\Delta m^2$ in $eV^2$ and the 
second is the value of $\psi_{\mu}$.  For $\psi$=0 there is no direct flavor
violation, so the ``source'' and ``detector'' labels are not meaningful.
$P_{\mu\rightarrow\tau}$=0.0006, appropriate for NOMAD.}
\end{figure}
The most noteworthy feature of Fig. 6 is that there is a ``confusion effect''.
The boundary curves for $\Delta m^2$=3 and 6 $eV^2$ for different flavor
violation situations crowd together and give a portrayal of the complications
that arise when direct flavor violations and oscillations become competitive.
The smaller values of $\psi$ used in the ``source-plus-detector'' situation
compared to the ``source-only'' are forced by the lack of solutions to the
probability equation when $|\psi|$ is 0.02.  The reason is that there is
a leading term $4\times tan^2 \psi$, so that there is no way to obtain
P=0.0006 when the lead term is 0.0016 and the $sin^2 x$ term is as small as
it is in the NOMAD experiment because of the small L/E.

\subsection{$\mu$-Decay as the Neutrino Source}

With muon decay as the neutrino source, there are electron and muon neutrino
appearance examples within this two-state system involving the
flavor-violating coefficients $F_{V2111},F_{V2212}$ and $F_{V2112}$. 
We give general expressions for cases where $\nu_{e}\rightarrow\nu_{\mu},
\nu_{e}\rightarrow\nu_{\tau},\nu_{\mu\rightarrow e}$, and $\nu_{\mu}
\rightarrow\nu_{\tau}$ in Appendix A.
Again restricting ourselves to a V-A form for the new interactions, the
general form of the probability factor for $\mu\rightarrow e$ transition is 

\begin{eqnarray}P_{\mu\rightarrow e}=&|-2isinx e^{-ix}sin\theta cos\theta
 F_{V2211}+(1-2isinx e^{-ix}sin^2 \theta) F_{V2111}|^2 + \nonumber \\
 &|-2isinx e^{-ix}sin\theta cos\theta F_{2212} +
(1-2isinx e^{-ix} sin^2 \theta) F_{V2112}|^2.\end{eqnarray}
The corresponding expression for $P_{e\rightarrow\mu}$ is obtained by the
interchanges
$F_{V2111}\leftrightarrow F_{V2212}$ and $cos\theta\leftrightarrow sin\theta$.
As in the $\pi$-decay case, Fig.(1), unless phases of the F's are of order 1
they play an insignificant role in the probability factors and we take them
to be zero in the following discussion.  Referring to Table 1, we see that
we may drop the $F_{V2111}$=$tan\psi_{ee}$ term since it is several orders of 
magnitude smaller
than the other lepton number violating coefficients in $\mu$-decay.  We take
$F_{V2211}$=1 and expand the resulting expressions using the parameterizations
$F_{V2112}$=$tan\psi_{e\mu}$ and $F_{V2212}$=$tan\psi_{\mu\mu}$.  The 
resulting expressions are given by

\begin{equation}P_{\mu\rightarrow e}=tan^2 \psi_{e\mu}+sin^2 x sin^2 2\theta
(1+tan^2\psi_{\mu\mu}-tan^2 \psi_{e\mu}-2cot 2\theta tan\psi_{\mu\mu}tan
\psi_{e\mu}) \end{equation}
and

\begin{equation}P_{e\rightarrow\mu}=tan^2 \psi_{\mu\mu}+(sin^2 x sin 2\theta
sin 2(\theta+\psi_{\mu\mu})) cos^{-2} \psi_{\mu\mu} + tan^2 \psi_{e\mu} (
1-sin^2 x sin^2 2\theta)).\end{equation}
Dropping the $\psi_{ee}$ in the expressions for $P_{\mu\rightarrow e}$ and 
$P_{e\rightarrow\mu}$
produces the lack of symmetry between the two equations.

It is clear from Eq.(27) and the restriction $|F_{V2112}|\equiv
|tan\psi_{e\mu}|\leq$ 8$\times$0.003 =0.024 that the effect of direct
interactions
on the amplitude of $\mu\rightarrow e$ oscillation will be important only if 
it turns out that $P_{\mu\rightarrow e}\leq$ 5x$10^{-4}$ and experiments can
explore that region\cite{Rub}. We return to this issue below.
The $e\rightarrow \mu$ oscillation amplitude can be more strongly affected,
because the branching fraction for $\mu\rightarrow e+\nu_{\mu}+\overline
{\nu_{\mu}}$ is not directly constrained by bounds on isospin related, purely 
charged-lepton processes.   Therefore, on these grounds alone, sizeable
effects cannot be excluded in the $\nu_{e}\rightarrow \nu_{\mu}$ appearance
case.\footnote{In specific models, indirect constraints on $F_{V2212}=
tan\psi_{\mu\mu}$ from $\mu\rightarrow e\gamma$ limits apply to a combination
of amplitudes that includes a $tan\psi_{\mu\mu}$.}
The L/E dependence of $P_{e\rightarrow\mu}$ in Fig. 7 demonstrates
the strong effect that the $\psi$ value has on the amplitude.

\begin{figure}[t,b]\hbox{\hspace{6em}\hbox{\psfig{file=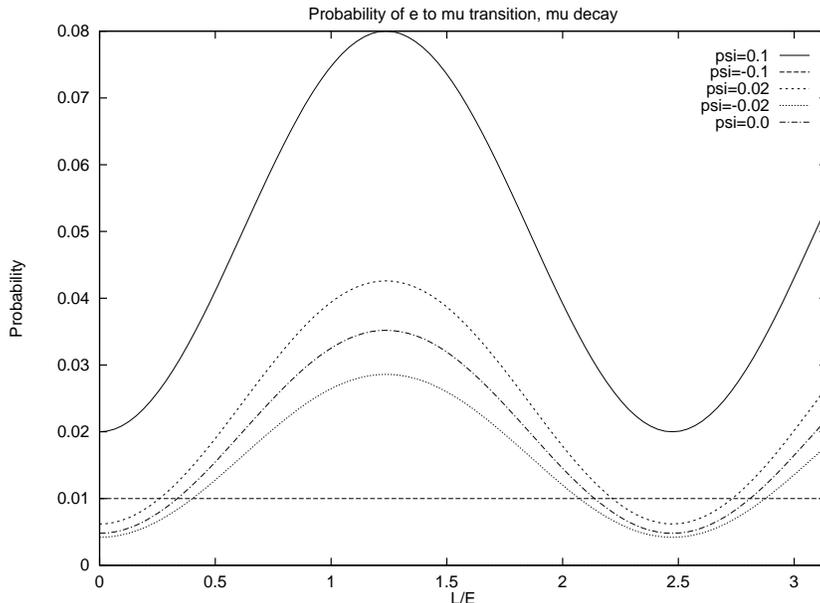,height=
8 cm,angle=-90}}}\caption{Dependence of $P_{e\rightarrow \mu}$ for a
$\mu$-decay source on L/E
for $\theta=0.1$, $\psi_{e\mu}=0.0$ and several different values of
$\psi_{\mu\mu}$ and with $\Delta m^2 = 1eV^2$.}\end{figure}
Here we show the oscillation probability as a function of L/E for several
choices of $\psi_{\mu\mu}$ with $\psi_{e\mu}$=0 and $\theta$=0.1. 
The effects of $\psi_{e\mu}$ are negligible with the chosen values of
$\psi_{\mu\mu}$ and $\theta$ 
in this case also, as one anticipates from inspection of Eq.(29) and the
bound on $tan\psi_{e\mu}$ mentioned above.  Note that the factor
$sin2(\theta+\psi_{\mu\mu})$ in Eq.(29) kills the oscillations in the
case $\psi$=-0.1, simply because the choice $\theta$=0.01 is made for the 
graph.
  
\subsubsection{Comparison of $\nu_{e}$ Appearance to $\nu_{\mu}$ 
Appearance}

The behavior of $P_{e\rightarrow\mu}$ shown in Fig. 7 can translate into
significant differences in the appearance probability for e compared to 
$\mu$ in neutrino experiments whose beams are extracted from muon collider
storage rings, for example.  In Fig. 8 we show the ratio of
$P_{\mu\rightarrow e}$ to $P_{e\rightarrow\mu}$ as a function of
$\psi_{\mu\mu}$, with $\theta$=0.003 and $\Delta m^2$=$1eV^2$, chosen to
be in a range allowed by LSND and not excluded by other experiments. 
  
\begin{figure}[t,b]\hbox{\hspace{6em}\hbox{\psfig{file=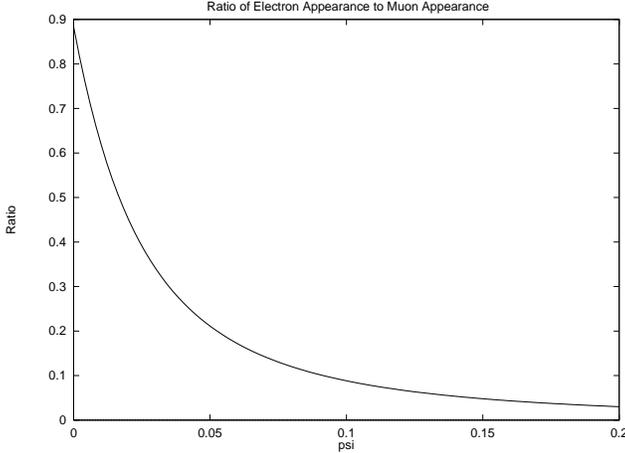,height=
6 cm,angle=-90}}}\caption{Dependence of the ratio $P_{\mu\rightarrow e}$/
$P_{e\rightarrow \mu}$ for a $\mu$-
decay source on $\psi_{\mu\mu}$ for $sin^2 2\theta$=0.003, 
and with $\Delta m^2 = 1ev^2$.  See text for treatment of L/E.}\end{figure}
The parameters chosen for Fig. 8 are guided by a recent proposal for
a medium baseline appearance search with $\nu$ beams from $\mu$ decay
\cite{Rub}. Below we summarize the relevant parameter values.
\medskip
\begin{center}
\begin{tabular}{|l|l|l|r|}                      \hline                        
                
\emph{$sin^2 2\theta$=0}&\emph{$\langle L/E\rangle$} &\emph{$\sigma$}&\emph{$sin^2 2\theta$=1}\\ \hline
 $\nu_{\mu}$ events=11600 $\mu^-$ & 0.66& 0.20 & $\nu_{e}$ events=6400 $e^-$ \\
 $\overline{\nu_{e}}$ events= 5070 $e^+$ & 0.75& 0.26&$\overline{\nu_{\mu}}$ events=3280 $\mu^+$\\ \hline 
\end{tabular}
\end{center}
\medskip
As Fig. 8 shows, the value of the ratio changes rapidly
as a function of $\psi$, and offers a possible method to directly constrain
$\psi$ down to 0.01 or less by comparing the $\nu_{\mu}\rightarrow\nu_{e}$
oscillation signal to its inverse.  We can put this another
way by comparing estimates of the number of $e^-$ events with the number of 
$\mu^+$ events detected downstream from $\mu^-$'s decaying in flight.
In a purely oscillation picture, the $e^-$
result from $\nu_{\mu}$ oscillating to $\nu_{e}$, while the $\mu^+$
result from $\overline{\nu_{e}}$ oscillating to $\overline{\nu_{\mu}}$.
In Fig. 9  we show a plot of the number of events expected vs. $\psi$ for the
$sin^2 2\theta$ and $\Delta m^2$ values assumed, given the SM event rates
estimated in the search experiment proposed in\cite{Rub}.

\begin{figure}[t,b]\hbox{\hspace{6em}\hbox{\psfig{file=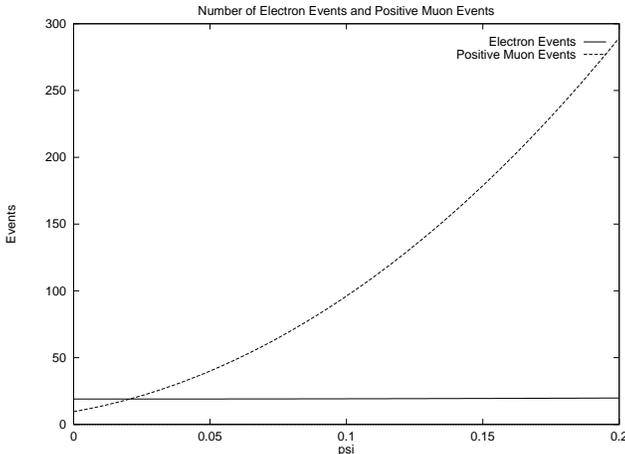,height=
6 cm,angle=-90}}}\caption{The numbers of $e^-$ and $\mu^+$ appearance events
as a function of $\psi_{\mu\mu}$, $\psi_{e\mu}$=0, for $sin^2 2\theta$=0.003, 
and with $\Delta m^2 = 1eV^2$. See text for treatment of L/E.}\end{figure}
Figure 9. makes evident that interference between the oscillation and 
direct amplitudes greatly enhances the $\mu^+$ appearance rate estimate
as $\psi$ grows, while there is little effect (a factor $sec^2\psi_{\mu\mu}$)
on the $e^-$ rate.  The numbers for $\psi$=0 correspond to the pure oscillation
numbers given in\cite{Rub}, but decreased by the factor $sin^2 2\theta$=
0.003 for $\Delta m^2$=1$ev^2$, since the numbers there refer to 
$sin^2 2\theta$=1.  We see again that looking for new physics may be
quite fruitful in the comparisons of appearance signals in
``pure'' neutrino beams provided by muon colliders. The current $\emph{direct}$
limit on $\psi_{\mu\mu}$ can be extracted from the limit \cite{berg83}  
$\sigma(\overline{\nu_{\mu}}+e^-\rightarrow \mu^- +\overline{
\nu})/\sigma(\nu_{\mu}+e^-\rightarrow\mu^-+\nu)=tan^2\psi$$ \leq 0.05$, or
$tan\psi \leq$ 0.22.
As Figs. 8  and 9  indicate, finding $N_{e^-}\geq N_{\mu^+}$ is sufficient
to improve the bound roughly to $\psi\leq 0.02$, an order of magnitude better
than the bound inferred from \cite{berg83}.  Allowing $|\psi_{e\mu}|$= 0.025
has little effect on this statement.

\subsubsection{Impact on Boundaries in e-$\mu$ Mixing Space}

Referring next to Eq.(28), let us consider what a proposed reach to $P_{\mu
\rightarrow e}\leq3.5\times10^{-4}$ at $90\%$ C.L.\cite{Rub} affects the
impact of direct interactions on the $\Delta m^2$ vs. $sin^2 2\theta$ contour.
Specifically, how do $\mu^-\rightarrow e^-\nu_{\mu}\overline{\nu_{\mu}}$
and $\rightarrow e^-\nu_{e}\overline{\nu_{\mu}}$ sources affect $\nu_{e}$ 
appearance from $\mu^-$ sources (or $\overline{\nu_{e}}$ appearance from
$\mu^+$ sources).  From (28) one can show that $tan^2 \psi_{e\mu}\leq
P_{\mu\rightarrow e}/(1-tan^2\psi_{\mu\mu}sin^2 x)$, which essentially means
that $tan^2\psi_{e\mu}\leq P_{\mu\rightarrow e}$ for small $tan\psi_{\mu\mu}$.
In particular for $|tan\psi_{\mu\mu}|\leq 0.2$ and $P_{\mu\rightarrow e}$
=$3.5\times10^{-4}$, $0.0187\leq (tan\psi_{e\mu})_{max}\leq 0.0193$. This
range is slightly less than the model-independent estimate, albeit on the
generous side, of the bound $|\psi_{e\mu}|\leq0.024$  that follows from the
experimental limit on muonium-antimuonium transition\cite{mumu}. 

We show the $P_{\mu\rightarrow e}=3.5\times10^{-4}$ contour in Fig. 10 with
several sets of values of $\psi_{\mu\mu}$ and $\psi_{e\mu}$.

\begin{figure}[t,b]\hbox{\hspace{6em}\hbox{\psfig{file=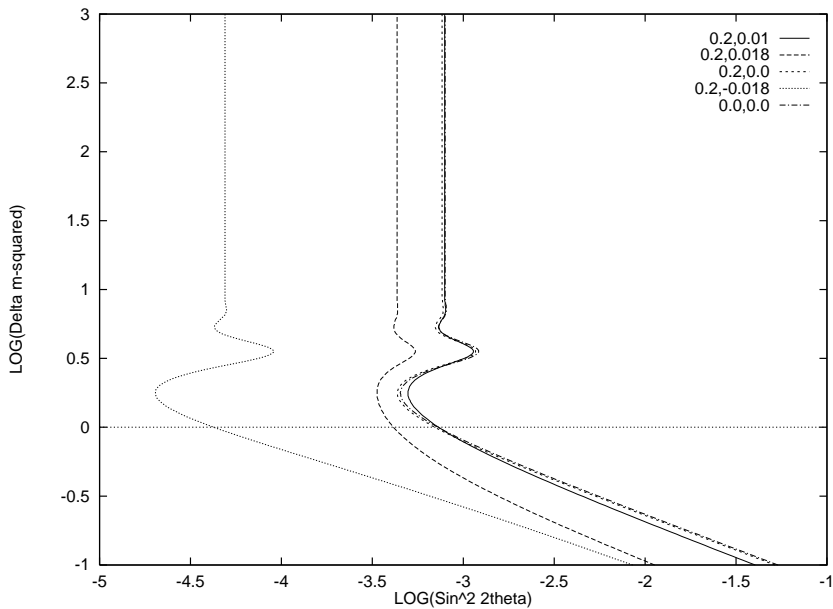,height=
8 cm,angle=-90}}}\caption{$P_{\mu\rightarrow e}$=$3.5\times10^{-4}$
boundaries at $90\%$ C.L. for several sets of $\psi_{\mu\mu}$, 
$\psi_{e\mu}$ values as shown on the legend.}\end{figure}The lack of
sensitivity to $\psi_{\mu\mu}$ with $\psi_{e\mu}\simeq0$, already
evident from Eq.(28) and Fig. 9 is indicated in Fig. 10 by the pile-up of
curves with $\psi_{e\mu}\leq 0.01$.  The situation changes drastically when
$\psi=0.018$, near its upper bound for solutions for $P=3.5\times10^{-4}$ 
to exist for some $\Delta m^2, \theta$ range of values. 
As noted in the introduction, possible direct
interaction contributions are too small to affect the LSND oscillation 
signal fits.  If LSND results are not confirmed and the limits drop to the
level of $P_{\mu\rightarrow e}\approx 10^{-4}$, then the interpretation of
those bounds should include the possible range of direct interaction strengths
allowed by limits from other experiments.  Correlated limits on $\Delta m^2$,
$sin^2 2\theta$ and $tan\theta_{ll'}$ can then be studied with the new data.

\section{Summary of Results}

Direct lepton number violating interactions and neutrino mass terms often go
hand-in-hand in physics beyond the standard model.  With this in mind
we developed a general parametrization of lepton number violating interactions
that fits smoothly with the usual description of neutrino oscillations in
terms of mixing angles.  We defined generalized ``probability factors''  and
illustrated with a number of examples drawn from accelerator appearance
experimental set-ups.  We concentrated on the case where the structure of the
effective four-fermion, charged current, flavor-violating operators is V-A.
This is the cleanest situation and lends itself to a parameterization in
terms of angles that makes the role of new interactions in the ``wrong-flavor''
appearance probability factors rather transparent.  For example, in Eq.(23)
the direct effects are expressed in terms of a leading term that gives all
of the appearance probability when there is no oscillation, a second term
that gives the interplay between the oscillation and direct effects and shows
that there is a special condition $\theta$-$\psi$ =integer$\times\pi/2$
where the usual oscillation term can be zero even
when there is a neutrino mass difference and a mixing between flavors.
The role of the CP-violating phase in the flavor-violating coupling is
isolated in the last term, which is zero when the phase is a multiple of
$\pi$, so the entire expression can be decomposed into readily
interpretable pieces.

Because the limits on lepton flavor violation are generally tight, the 
impact on the oscillation picture of appearance and disappearance 
is limited to cases where the oscillation probabilities are small and
sensitive to the precise values of the mixing angles. We illustrate
such situations in detail in Sec. 4, where the sensitive dependence of
the appearance probability on the relative values of the flavor-violating
angle $\psi$ and the mass-mixing angle $\theta$ are shown in Figs. 1, 2 and
7 for the amplitude of the oscillation behavior, are shown in Fig. 3 for the
the NOMAD, $\mu\rightarrow\tau$, two-flavor mixing boundaries and are shown
in Fig. 6 for the corresponding three-flavor mixing boundaries in the
one-mass-difference dominance model. The NOMAD experiment examples
are carefully done within the limits imposed on the size of the flavor
violations, as summarized in Table 1.  We present similar considerations
for the case where the source is $\mu$-decay, and the boundaries appropriate
to a proposed $\mu$ storage ring environment are shown in Fig. 10, again
for realistic constraints on the flavor-violation strength.  

Fig. 4 illustrates another new effect
that shows that care must be taken in interpreting limits set by wrong-flavor
appearance searches. The behavior of the $P_{\mu\rightarrow e}$=$10^{-4}$
boundaries for source-only and source-plus-detector are quite different as
viewed in the $\theta$-$\psi$ plane.  The former boundary is ``tilted''
and so the bound on $|\psi|$ set by the $90\%$ C.L. boundary is correlated
with the value of $\theta$ so that, contrary to the usual expectation, the
bound on $\psi$ is not obtained by setting $\theta$=0.  The boundary for
the latter situation, however, does conform to the usual expectation.

We show how the comparison of electron appearance and muon appearance 
experiments in a clean $\mu$ storage ring environment could be used to
increase the sensitivity to direct, neutrino flavor violations by an order
of magnitude in the discussion of Figs 8 and 9.  These show the rather
strong sensitivity of the muon appearance effects on the value of the
violation parameter $\psi_{\mu\mu}$.

\section{Concluding Remarks}
A theme that recurs throught the analysis in Sec.4 is that
small, direct flavor violation can seriously complicate the picture of
wrong-flavor appearance in a number of experimentally realistic situations.
In short, the interpretation of a signal can be quite ambiguous.  To 
sort out the complete picture, a number of measurements at various values
of L/E in a variety of different channels is needed.  Perhaps the cleanest
and most flexible environment for such studies is provided by a $\mu^+$
$\mu^-$ collider with an associated facility for neutrino beams .  Studies
such as those exemplified in\cite{Rub} can put much more stringent tests
on direct neutrino flavor violation than currently exist, while at the same
time allowing detailed oscillation analysis. Direct flavor-violation effects in
semileptonic processes require pion and kaon beams of course, and the 
BooNE experiment will provide exploration of a wide range of parameter space,
for example.

Though the direct flavor-violating strengths are constrained to be small, 
the matter-enhancement effects can lead to large transition probabilities,
as analyzed in \cite{GG} and \cite{forn}, and for small probabilities the
``vacuum'' effects in accelerator experiments can lead to important
modification of the oscillation description.  We conclude that the crucial
role neutrinos play in our understanding of particle and astrophysics
requires that data be analyzed with the relevant flavor violation included 
to properly interpret current and future experiments.
The present work provides a framework for that task and provides vivid, 
realistic examples of its application. 

\medskip

Acknowledgements: We thank Tom Weiler for discussions and John
Ralston for suggestions on the manuscript.  This work was supported
in part by U.S. DOE Grant No. DE-FGO2-85ER40214.  L.M.J. thanks the 
Department of Physics at Drury University for support during the course of
this work.

\appendix
\section{Muon Decay Appearance Probabilities}

We gather the general formulas for appearance probabilities
that apply when muon decay is the source of neutrinos.
The presence of two neutrinos in the final state,
only one of which is detected by a choice of lepton-flavor sensitive
detector, makes the analysis slightly different from the situation
when meson decay provides the source of neutrinos.
 
\begin{equation}P_{e\rightarrow\mu}=\sum_{k}|F_{2k1j}U_{jc}^* e^{-iE_{c}t}
 U_{1c}|^2. \end{equation}

\begin{equation}P_{\mu\rightarrow e}=\sum_{k}|F_{2j1k}^* U_{jc}
 e^{-E_{c}t}U_{1c}|^2. \end{equation}

\begin{equation}P_{e\rightarrow\tau}=\sum_{k}|F_{2j1k} U_{jc}^*
 e^{-iE_{c}t} U_{3c}|^2. \end{equation}

\begin{equation}P_{\mu\rightarrow\tau}=\sum_{k}|F_{2j1k}^* U_{jc}
 e^{-iE_{c}t} U_{3c}|^2. \end{equation}
In all of the expressions, the repeated indices within the absolute-square
are understood to be summed.  By choosing a given process,
keeping the dominant flavor-transition terms and parameterizing the
F amplitudes by angles, the $\mu$ decay source expressions in the
text can be reproduced from these expressions.

\end{document}